\documentclass[12pt]{spieman}  
\usepackage{amsmath,amsfonts,amssymb}
\usepackage{graphicx}
\usepackage{setspace}
\usepackage{tocloft}
\usepackage{blkarray}
\usepackage[english]{babel}

\title{Single-shot characterization of photon indistinguishability with dielectric metasurfaces}

\author[1,2,$\dagger$,*]{Jihua Zhang}
\author[1,$\dagger$]{Jinyong Ma}
\author[1]{Neuton Li}
\author[1,3]{Shaun Lung}
\author[1,*]{Andrey~A.~Sukhorukov}

\affil[1]{ARC Centre of Excellence for Transformative Meta-Optical Systems (TMOS), Department of Electronic Materials Engineering, Research School of Physics, The~Australian National University, Canberra, ACT 2600, Australia}
\affil[2]{Songshan Lake Materials Laboratory, Dongguan, Guangdong 523808, P. R. China}
\affil[3]{Abbe Center of Photonics, Friedrich-Schiller Universit{\"a}t, Albert-Einstein-Stra{\ss}e 15, 07745 Jena, Germany}
\affil[$\dagger$]{Equal contribution}
\affil[*]{Corresponding authors: \linkable{zhangjihua@sslab.org.cn}, \linkable{andrey.sukhorukov@anu.edu.au} }

\cftpagenumbersoff{figure}
\cftpagenumbersoff{table} 
\begin{document} 
\maketitle

\begin{abstract}
  Characterizing the indistinguishability of photons is a key task in quantum photonics, underpinning the tuning and stabilization of the photon sources and thereby increasing the accuracy of quantum operations. The protocols for measuring the degree of indistinguishability conventionally require photon-coincidence measurements at several different time or phase delays, which is a fundamental bottleneck towards the fast measurements and real-time monitoring of indistinguishability. Here, we develop a static dielectric metasurface grating without any reconfigurable elements that realizes a tailored multiport transformation in the free-space configuration without the need for phase locking and enables single-shot characterization of the indistinguishability between two photons in multiple degrees of freedom including time, spectrum, spatial modes, and polarization. Topology optimization is employed to design a silicon metasurface with polarization independence, high transmission, and high tolerance to measurement noise. We fabricate the metasurface and experimentally quantify the indistinguishability of photons in the time domain with fidelity over 98.4\%. We anticipate that the developed framework based on ultrathin metasurfaces can be further extended for multi-photon states and additional degrees of freedom associated with spatial modalities. 
\end{abstract}



\begin{spacing}{2}   

\section{Introduction}
%
The indistinguishability between photons is strongly related to quantum interference and coherence~\cite{Mandel:1991-1882:OL},
and it plays a critical role in photon-based quantum information technologies~\cite{Lal:2022-21701:AVSQ}. For example, indistinguishable photons are essential resources for photonic quantum computing, as their interference 
underpins the implementation of two-photon logic gates~\cite{Knill:2001-46:NAT}. 
When the photons are not indistinguishable, it creates ``mode-mismatch'' errors in optical quantum computation and simulation algorithms~\cite{Rahimi-Keshari:2016-21039:PRX}. On the other hand, photon indistinguishability is also a prerequisite for long-distance quantum communications in terms of realizing quantum repeaters via entanglement swapping \cite{Duan:2001-413:NAT}.
Therefore, characterizing the degree of indistinguishability of photons 
is of great importance in many applications~\cite{Viggianiello:2018-1470:SCB, Brod:2019-63602:PRL, Schofield:2022-13037:PRR, Brunner:2018-210401:PRL, Pont:2022-31033:PRX}. 

The indistinguishability of two photons 
is conventionally characterized by the Hong-Ou-Mandel (HOM) interference experiment with a balanced beam splitter by comparing the two-photon coincidences between the two output ports at several different time delays~\cite{Hong:1987-2044:PRL}. An alternative approach is based on a Mach-Zehnder interferometer by measuring the output two-photon coincidences while scanning the phase delay in one arm~\cite{Rarity:1990-1348:PRL}. A recent work generalized the latter method to the multi-photon case through a cyclic integrated interferometer with a tunable phase delay~\cite{Pont:2022-31033:PRX}. In all these previous experiments, the indistinguishability is quantified by the visibility of the multi-photon-coincidence fringes,
thereby requiring a reference at the fringe plateau that is determined from several measurements at varying time or phase delays. These delays are typically implemented by mechanical tuning of bulky free-space optical elements or thermal and electrical tuning of integrated photonic circuits. 
Therefore, these approaches are not well suited for fast measurements of indistinguishability.
On the other hand, single-shot characterization of photon indistinguishability, if realized, could enable real-time monitoring, improve precision, and reduce the size and power consumption by removing the need for delay tuning. In particular, real-time monitoring of photon indistinguishability has practical applications in the development of fully indistinguishable 
photon sources by introducing a feedback control to tune the operation parameters of the source such that it reaches the peak indistinguishability and afterwards stabilizes at the peak~\cite{Sayrin:2011-73:NAT, Raghunathan:2009-33831:PRA}.

Metasurfaces containing a thin layer of subwavelength nanostructures allow for parallel measurements and full characterization of quantum states by encoding multiple transformations~\cite{Li:2020-100584:PRSS, Solntsev:2021-327:NPHOT, Liu:2021-200092:OEA, Wang:2022-38:PT, Ji:2023-169:LSA}. Two-photon interference and state reconstruction were recently demonstrated with metasurfaces~\cite{Stav:2018-1101:SCI, Wang:2018-1104:SCI, Georgi:2019-70:LSA, Lyons:2019-11801:PRA, Li:2021-267:NPHOT, Wang:2023-3921:NANL}, whereas the potential of metasurfaces for direct characterization of photon indistinguishability remained unexplored. 
In this work, we develop a static metasurface-enabled interferometer to characterize the two-photon indistinguishability in several degrees of freedom including time, spectrum, spatial modes, and polarization, without a need for multiple measurements at different time or phase delays. This enables not only real-time characterization but also eliminates extra measurement errors and minimizes the device size due to the no longer needed delay tunability. Topology optimization is employed to design a silicon metasurface with polarization independence, high transmission, and high tolerance to measurement noise. 
The designed metasurface essentially represents an ultra-stable interferometer supporting a tailored multiport transformation, which remained a challenge in the free-space setups due to the requirement of deep subwavelength stability and its implementation has typically relied on the integrated waveguide photonics circuits~\cite{Carolan:2015-711:SCI} or phase locking~\cite{Zhong:2020-1460:SCI}.
%
In an experiment, we fabricate the designed metasurface and perform the measurements for two photons with the same polarization and different levels of indistinguishability in the time degree of freedom, showing an agreement with the conventional HOM-type characterization based on delay tuning with a fidelity over~98.4\%. 

%
\section{Results and discussion}
\begin{figure*}[t]
    \centering
    \includegraphics[width=1\textwidth]{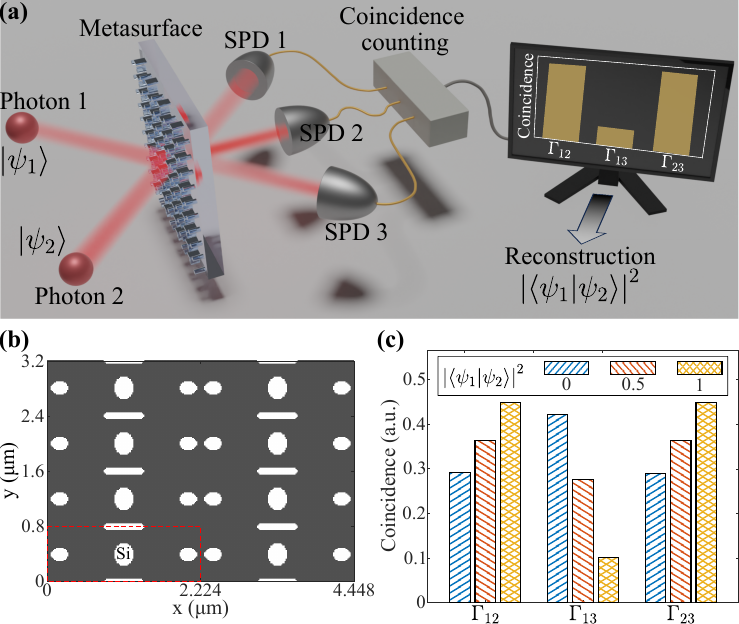}
    \caption{(a)~Schematic of the metasurface-enabled single-shot characterization of two-photon indistinguishability. Two photons with separable states $\psi_{1}$ and $\psi_{2}$ are input on the metasurface grating and diffracted into three directions in transmission. Simultaneous measurement of the three different two-photon coincidences at the output enables the realtime reconstruction of the two-photon indistinguishability which is quantified by $| \langle \psi_1 | \psi_2 \rangle |^2$. (b)~Silicon metasurface designed by topology optimization. The dashed red box marks out a unit cell with a dimension $2224$~nm$\times 800$~nm. (c)~Predicted output coincidence distributions for two incident photons which are fully distinguishable ($| \langle \psi_1 | \psi_2 \rangle |^2=0$), partially distinguishable ($| \langle \psi_1 | \psi_2 \rangle |^2=0.5$), and indistinguishable ($| \langle \psi_1 | \psi_2 \rangle |^2=1$).}
    \label{fig:1}
\end{figure*}

Figure~\ref{fig:1}(a) illustrates the concept of the metasurface-enabled single-shot characterization of the indistinguishability between two photons that are injected 
from different incident angles and interfere with each other along three output paths. Essentially, the metasurface functions as a multiport interferometer with two inputs and three outputs. We consider two photons with separate wavefunctions defined as $|\psi_1\rangle$ and $|\psi_2\rangle$. For the case where the detectors cannot distinguish the two photons by their internal properties such as spectra or polarization, the incident two-photon split state can be described by a reduced spatial density matrix in the form~\cite{Zhang:2023-62615:PRA}:
%
\begin{equation} \label{eq:densitymatrix}
\hat{\rho} = 
\begin{blockarray}{ccc}
& \langle 12| & \langle 21| \\
\begin{block}{c[cc]}
  |12\rangle & \rho_1 & \rho_2 \\
  |21\rangle & \rho_2 & \rho_1 \\
\end{block}
\end{blockarray} ,
\end{equation}
where all the other elements are zero. 
The matrix is defined by only two free parameters $\rho_1=0.5\langle\psi_1|\psi_1\rangle \langle\psi_2|\psi_2\rangle$ and $\rho_2=0.5\langle\psi_1|\psi_2\rangle \langle\psi_2|\psi_1\rangle$. 
We take into account the particular form of the density matrix to develop a more efficient measurement approach compared to the general quantum tomography methods~\cite{James:2001-52312:PRA, Wang:2018-1104:SCI, Czerwinski:2022-268:OPT-MDPI}.

The two-photon indistinguishability is quantified by
\begin{equation} \label{eq:dfunction}
    I_{2p} = \frac{\rho_2}{\rho_1} = \frac{|\langle \psi_1 | \psi_2 \rangle|^2}{\langle\psi_1|\psi_1\rangle \langle\psi_2|\psi_2\rangle} .
\end{equation}
In the conventional HOM experiment, the two-photon coincidence at each time delay, normalized to the plateau at large time delays, would be equal to $1-I_{2p}$. Accordingly, the depth of the HOM dip is related to the indistinguishability of two photons at zero time delay. Specifically, $I_{2p}=0$, $I_{2p}=1$, and $0<I_{2p}<1$ represent the cases where the two photons are distinguishable, indistinguishable, and partially distinguishable, respectively. 
Notably, the range of $I_{2p}>0.5$ is commonly considered as a quantum signature~\cite{Kim:2013-63843:PRA}, since such visibility could only be achieved classically by specially preparing pulses with particular phase differences~\cite{Sadana:2019-13839:PRA}.

We can determine the value of $I_{2p}$ by measuring the two-photon coincidences between pairs of distinct output ports, which are related to the input density matrix elements as 
\begin{equation}\label{eq:correl_ij}
    \Gamma_{ij}=(T_i \otimes T_j)\hat{\rho}(T_i^\dag \otimes T_j^\dag) .
\end{equation}
Here $T_i$ is the $i$-th row of the $3\times2$ metasurface transmission matrix $\textbf{T}=[t_{11},t_{12};t_{21},t_{22};t_{31},t_{32}]$ and $\dag$ denotes complex conjugation. Then, the output two-photon coincidences are determined by the nonzero density matrix elements of the input two-photon state as
\begin{equation} \label{eq:correl}
\begin{bmatrix}
\Gamma_{12}\\
\Gamma_{13} \\
\Gamma_{23}
\end{bmatrix} =
\mathbf{M}
\begin{bmatrix}
\rho_1\\
\rho_2
\end{bmatrix} .
\end{equation}
Here $\textbf{M}$ is a real-valued $3\times2$ matrix that is determined by the metasurface transmission matrix $\textbf{T}$ through Eqs.~(\ref{eq:densitymatrix},\ref{eq:correl_ij}). Therefore, after characterizing $\textbf{M}$ and measuring the output coincidences, the input density matrix elements can be directly reconstructed via $[\rho_1~\rho_2]^T=\textbf{M}^{+}[\Gamma_{12}~\Gamma_{13}~\Gamma_{23}]^T$, where $\textbf{M}^{+}$ is the pseudoinverse of $\textbf{M}$. However, the density matrix obtained by this method might be nonphysical in the presence of experimental errors. Therefore, we perform the reconstruction via the maximum likelihood estimation by finding a physical density matrix that results in a coincidence distribution closest to the measured one~\cite{James:2001-52312:PRA}. Specifically, we impose the constraint $\rho_1 \geq \rho_2 \geq 0$ for the density matrix to be physically possible. Finally, we obtain $I_{2p}$ using Eq.~(\ref{eq:dfunction}). 

The next research question lies in how to design the metasurface with an optimized $\textbf{T}$ to implement such a reconstruction. To simplify the design, we consider a symmetric metasurface with $t_{32}=t_{11}$, $t_{22}=t_{21}$, and $t_{12}=t_{31}$. 
In this case, the explicit form of $\textbf{M}$ is
\begin{equation}
    \textbf{M} = \begin{bmatrix}
2(|t_{11}|^2+|t_{31}|^2)|t_{21}|^2 & 4|t_{11}||t_{31}||t_{21}|^2\cos\delta \\
2(|t_{11}|^4+|t_{31}|^4)  &  4|t_{11}|^2|t_{31}|^2\cos2\delta \\
2(|t_{11}|^2+|t_{31}|^2)|t_{21}|^2 & 4|t_{11}||t_{31}||t_{21}|^2\cos\delta
\end{bmatrix}
\end{equation}
where $\delta = \text{arg}(t_{31})-\text{arg}(t_{11})$.
We establish that in order to characterize with a single-shot measurement the two-photon indistinguishability in all degrees of freedom, including polarization, the transmission coefficients of the symmetric meta-grating should satisfy four conditions: (1)~$|t_{11}^{TE}|=|t_{11}^{TM}|$, (2)~$|t_{21}^{TE}|=|t_{21}^{TM}|$, (3)~$|t_{31}^{TE}|=|t_{31}^{TM}|$, (4)~$\delta^{TE}=\delta^{TM}$, where TE denotes horizontal and TM~-- vertical input polarization (more details are provided in the Supplementary~S1). In essence, these conditions require that the transmission is polarization-independent, up to a global phase. We further impose two practical requirements on the meta-grating: high total transmittance (thus low photon loss) for any polarization and robustness in the presence of measurement inaccuracies. Specifically, we aim to reduce the amplification of measurement noise during the reconstruction process~\cite{Wang:2018-1104:SCI}, which corresponds to minimizing the condition number of the matrix $\mathbf{M}$, defined as the ratio of its maximum and minimum singular values.

We use the topology optimization~\cite{Fan:2020-196:MRSB} approach to numerically design free-form metasurfaces. The best design maximizes the 
figure of merit (FOM), which we formulate taking into account all the  requirements, 
\begin{equation}\label{eq:fom}
   FOM =(F_{p})^6 \cdot \min \left[{\rm svd} \left(\textbf{M}^{TE} \right) \right] \cdot \min \left[{\rm svd} \left(\textbf{M}^{TM} \right) \right] .
\end{equation}
Here the minimum singular values (svd) of the system matrix reflect the total transmission efficiency and condition numbers and
\begin{equation}
    F_p = 
    \prod_{n = 1,2,3}
        \frac{4|t_{n,1}^{TE}||t_{n,1}^{TM}|}{(|t_{n,1}^{TE}|+|t_{n,1}^{TM}|)^2} \cdot 
    \left| \cos\left(\frac{\delta^{TE}-\delta^{TM}}{2}\right)\right|
\end{equation}
quantifies the similarity of two polarization transmissions up to a global phase.
We develop simulations using the MetaNet codebase~\cite{Jiang:2020-13670:OE} in combination with RETICOLO rigorous coupled wave analysis~\cite{Hugonin:2101.00901:ARXIV}.
Although, in contrast to previous topology optimization studies, the FOM in Eq.~(\ref{eq:fom}) has no simple analytical expression for the partial derivatives with respect to the transmission coefficients, we simply calculate them numerically.
Figure~\ref{fig:1}(b) shows the final meta-grating design for a nanopatterned 1000~nm thick silicon layer on a sapphire substrate, where the dimension of a unit cell is chosen as $a_x=2224$~nm and $a_y=800$~nm to enable three diffraction orders along the \textit{x}- and no high-order diffraction along the \textit{y}-direction respectively. As shown in Supplementary~S3, this metasurface achieves a high FOM at the target telecommunication band near the central wavelength (1560~nm) of the considered photon-pair source. Specifically, for TE and TM polarizations, the amplitudes and phases of transmission coefficients are very similar, and the condition number is 
close to the theoretically optimal performance of an integrated waveguide circuit in two-photon state tomography~\cite{Zhang:2023-62615:PRA}, and the total transmission reaches 90\% within the wavelength range 1550-1570 nm. More details on the structure and operation principle of the final metasurface are provided in Supplementary~S4.

We calculate the two-photon coincidence distributions at the outputs for different input states using Eq.~(\ref{eq:correl}) and the simulated optical response of the designed metasurface. Figure~\ref{fig:1}(c) shows the predicted output two-photon coincidence distributions for two completely distinguishable, partially distinguishable, and completely indistinguishable photons. The clear difference of coincidences between such states enables robust reconstruction of the input density matrix $\hat{\rho}$ and thus the two-photon indistinguishability $I_{2p}$. 

\begin{figure}[!t]
    \centering
    \includegraphics[width=1\columnwidth]{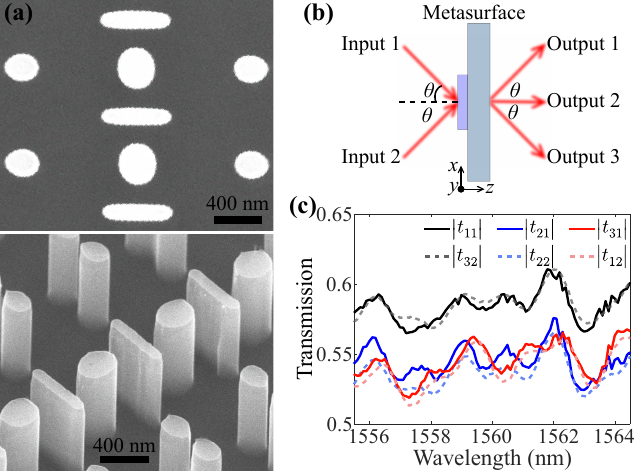}
    \caption{(a)~Top- and tilted-view scanning electron microscope images of the fabricated metsurface. (b)~Schematic of characterizing the transmission amplitudes from two input directions using a tunable laser. The incident angle was set to $\theta=\arcsin(\lambda_0/a_x)$ where $\lambda_0=1560$~nm is the center wavelength of the photons used in the next quantum experiment and $a_x=2224$~nm is the period of the metasurface along \textit{x} direction. (c)~Measured transmission amplitudes in a wavelength range near $\lambda_0$.}
    \label{fig:2}
\end{figure}
\begin{figure*}[!thb]
    \centering
    \includegraphics[width=1\columnwidth]{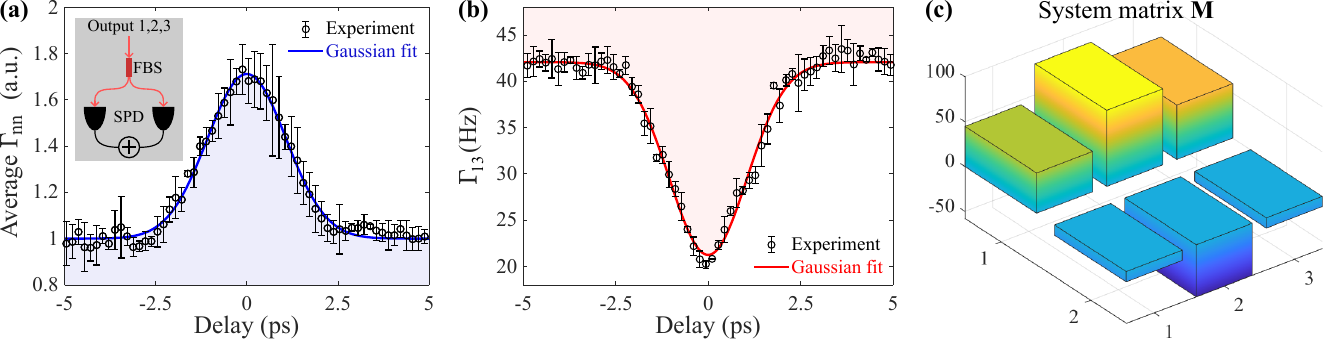}
    \caption{(a) Measured normalized two-photon coincidences at the same output port by coupling into a 50:50 fiber beam splitter (FBS) followed by two SPDs (see inset). The average from three outputs is plotted here. The calibrated $I_{2p}$ is shown on the right axis after a Gaussian fitting of the data. (b) Measured two-photon coincidences between output ports 1 and 3. For all data points in (a) and (b), five measurements were taken to obtain the average and error bars, which represent one standard deviation. (c) The characterized system matrix $\textbf{M}$ connecting the input density matrix and output coincidences.}
    \label{fig:3}
\end{figure*}

We fabricated the metasurface according to the optimal design, see the scanning electron microscope image in Fig.~\ref{fig:2}(a). We first characterize the metasurface using a classical setup to determine the transmission amplitudes as sketched in Fig.~\ref{fig:2}(b). A tunable continuous laser is used as the source and it inputs in the $y-z$ plane. Figure~\ref{fig:2}(c) shows the measured results for the TE polarization. The total transmission efficiency for two incident angles is up to 90\% at the designed wavelength. 
One can see that the transmission is close to a symmetric response as designed. 
Importantly, the proposed scheme is robust to fabrication imperfections, and in particular, the slight asymmetry has no influence on the operation. Indeed, we only need to perform a once-off characterization of the system matrix $\textbf{M}$ which then incorporates all the features of the actual fabricated metasurface, and accordingly allows the subsequent accurate reconstructions of the quantum two-photon density matrices. 
Analogous results were also obtained for the TM polarization (see Supplementary~S5). However, we found that the fabricated metasurface has a low similarity between the two polarizations. 
The possible reasons include the polarization dependent transmission at the substrate-air interface and fabrication deviation of the metasurface from the designed one. These can be mitigated by anti-reflection coatings on the back side of the substrate and further optimization of the fabrication recipes. The polarization dependence of the fabricated metasurface prevented us from characterizing the two-photon indistinguishability in the polarization degree of freedom. Therefore, in the following quantum measurements, we will fix the polarization to TE and characterize the two-photon indistinguishability in the time/spectrum domains. 
We also note that the classical experiment revealed random phase accumulations along the two beam paths before the metasurface, such that the phase difference between $t_{31}$ and $t_{11}$ could not be determined. However, we do not require this measurement since the matrix $\textbf{M}$ is calibrated directly based on quantum correlations as we describe below. On the other hand, the presence of such random phase fluctuations indicates that in our setup, the values of $I_{2p}>0.5$ can indeed serve as a quantum signature~\cite{Kim:2013-63843:PRA, Sadana:2019-13839:PRA}.


In the quantum experiment, as schematically shown in Fig.~\ref{fig:1}(a), two photons are generated from a PPKTP nonlinear crystal, and the power meters are replaced by single photon detectors (SPDs). A time delay is introduced between two photons to tune the degree of indistinguishability, resulting in 
\begin{equation} \label{eq:spec_overlap}
    \langle \psi_1 | \psi_2 \rangle = \int \phi_1^* (\omega) \phi_2 (\omega) e^{-i\omega \tau} d\omega
\end{equation}
where $\phi_i(\omega)$ is the spectrum of the $i$-th photon and $\tau$ is the delay between two photons.

In order to prove the validity and accuracy of our scheme, we first calibrate the $I_{2p}$ of the incident photon pair. As mentioned before, the conventional way to realize this is based on the HOM effect after a balanced beam splitter. Yet, an absolute calibration can be performed using the same metasurface without a need for other bulky optics.
Specifically, we 
implement a conventional Hanbury-Brown-Twiss experiment at each of the outputs 1,2,3 by adding a 50:50 fiber beam splitter followed by two SPDs and comparing the two-photon coincidences at different time delays, as shown by an inset in Fig.~\ref{fig:3}(a). Note that in such an experiment the two-photon-coincidences
are independent of the metasurface response after normalization to the plateau at large time delays and are solely determined by the spectrum similarity of two photons at each time delay, as in Eq.~(\ref{eq:spec_overlap}). Mathematically, the normalized two-photon coincidences at each time delay are related to the density matrix elements by $\Gamma_{nn}=1+\rho_2/\rho_1=1+I_{2p}$  (see Supplementary~S2), reaching a maximum value of 2 for zero delay between identical photons in agreement with Refs.~\cite{Fearn:1987-485:OC, Georgi:2019-70:LSA}. After collecting the experimental data (the two-photon source is described in Supplementary~S7), we average the normalized two-photon coincidences for the three outputs and perform a Gaussian fitting of the average data, both are plotted in Fig.~\ref{fig:3}(a). Based on this, we first determine the position of zero time delay at the peak position of the Gaussian fitting blue curve. Then, we calibrate the normalized density matrix and the related $I_{2p}$ of the incident photon pair from the Gaussian fitting data by considering $\rho_1=0.5$ and $\rho_2=\Gamma_{nn,{\rm fit}}/2-\rho_1$ at each time delay. We note that this essentially provides an efficient way to characterize the two-photon state and $I_{2p}$ without 
a beam splitter with a known transmission matrix to perform the HOM experiment. Of course, it still relies on the visibility of the interference fringe at different time delays to quantify the $I_{2p}$. The calibrated $I_{2p}$ on the right axis of Fig.~\ref{fig:3}(a) is 0.73 at zero time delay, consistent with the quantum regime of the photon pair source, and allowing us to validate the metasurface performance. 

We now proceed to the main demonstration of characterizing the $I_{2p}$ in a single shot by measuring the two-photon coincidences between pairs of distinct outputs. The measurement results are shown in Fig.~\ref{fig:3}(b) for $\Gamma_{13}$ and Supplementary Fig.~S9 for $\Gamma_{12}$ and $\Gamma_{23}$. Based on these measurements and the calibrated $\rho_{1,2}$ in the previous step, we are able to characterize the system matrix $\textbf{M}$ of the fabricated metasurface through a nonlinear fitting of the data using Eq.~(\ref{eq:correl}). The characterized $\textbf{M}$
is shown in Fig.~\ref{fig:3}(c), and the numerical values are presented in the Supplementary Eq.~(S17). It is close to a symmetric matrix as designed. The slight asymmetry could arise from the different fiber-coupling and detector efficiencies at outputs~1 and~3. Nevertheless, all these measurement deviations are accounted for in the characterization process, which makes our protocol highly robust to the detection performance. The corresponding condition number of the characterized $\textbf{M}$ is $\simeq 5.2$, 
which
allows us to reconstruct the input density matrix with low amplification of the measurement noise.

Finally, based on this characterized system matrix and the maximum likelihood estimation method~\cite{James:2001-52312:PRA}, we reconstruct the density matrix and two-photon indistinguishability at each time delay by the three two-photon coincidences at the same time-delay point. More details on the data processing and the reconstruction process are provided in Supplementary~S8. Figure~\ref{fig:4}(a) shows the reconstructed input density matrix elements and the calibrated input at each time delay. We see that the reconstructed values are close to the calibrated input, especially for the input states with a high degree of indistinguishability near the zero time delay. The average fidelity between the calibrated 
and reconstructed 
density matrices 
calculated using the formulation in Ref.~\cite{Jozsa:1994-2315:JMO} (see Supplementary Eq.~(S22))
is found to be as high as 99.88\%, with a minimum fidelity of 98.4\%. The characterized v.s. the calibrated input $I_{2p}$ is shown in Fig.~\ref{fig:4}(b), which confirms a good agreement, especially for photon pairs with $I_{2p}\geq 0.5$. Here the results for positive delays are presented, and a similar plot for negative delays is provided in Supplementary Fig.~S10. Since $I_{2p}$ directly relates to the HOM visibility, $I_{2p}\geq 0.5$ represents the quantum signature of two photons and cannot be mimicked by classical pulses without phase control~\cite{Sadana:2019-13839:PRA}. Different from HOM measurement, in this scheme the needed two-photon coincidences can be measured simultaneously at a fixed time delay, which enables the fast and efficient characterization of the indistinguishability of the incoming photon pairs. An additional benefit is that the measurement errors due to the no longer needed phase or delay tuning are eliminated. 

%
\begin{figure}
    \centering
    \includegraphics[width=1\columnwidth]{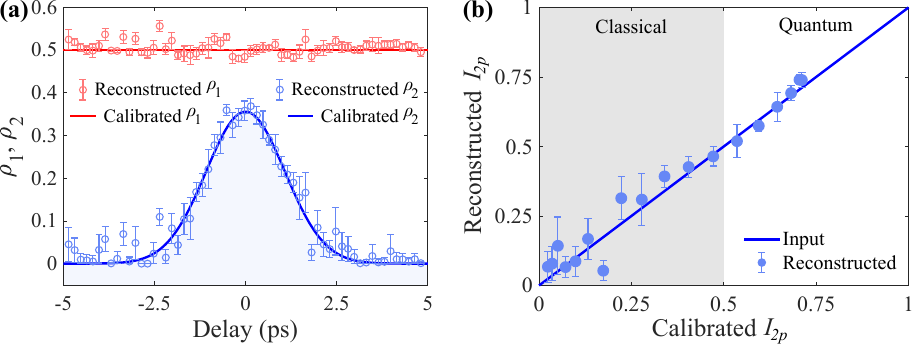}
    \caption{(a)~Reconstructed density matrix elements of the input two-photon states and the corresponding calibrated inputs at different time delays. (b)~Reconstructed two-photon indistinguishability v.s. the calibrated input ones at different positive time delays. The error bars represent one standard deviation. The solid line denotes a theoretically ideal reconstruction. White shading marks the region of $I_{2p}>0.5$ which cannot be reached by classical
    pulses without phase control. }
    \label{fig:4}
\end{figure}

\section{Conclusion}
To conclude, we have proposed and demonstrated a metasurface-based multiport interferometer to characterize the two-photon indistinguiability. Different from conventional methods requiring the measurement of two-photon coincidences at several time delays or using reconfigurable setups, the presented scheme uses a static metasurface and operates in a single shot. These features can facilitate realtime monitoring of photon indistinguishability with ultra-stable performance due to a realization of multiport phase-sensitive interference in a single integrated metasurface (see more discussions in Supplementary~S6) and the nonexistence of reconfiguration errors. 

It is noteworthy that the proposed scheme can still operate in real-time even when the photon flux rate varies since our protocol is based on the relative magnitude of the three output photon coincidences at the same time. In contrast, for the conventional HOM method, one would need to continuously recalibrate the coincidence counts for fully distinguishable photons at large time delays. Furthermore, the same metasurface can work not only for the TE polarization as we demonstrated in this work, but also for the TM polarization after a new calibration of the system matrix $\textbf{M}$. 
The metasurface interferometer can also allow characterization for photons coming from two angles different from those considered in this work by tuning the orientation and period size of the metasurface followed by a topological optimization to find the corresponding nanopattern.

The metasurface can also be designed for a larger number of output channels by increasing the unit cell size to allow for more diffraction orders. This enables more parallel coincidence measurements that may allow the developed principle to be 
further extended for multi-photon states and additional degrees of freedom associated with spatial modalities, providing versatile and ultracompact integrated quantum optical elements for various applications. In particular, the tiny size and ultra-light weight of the metasurface and zero power consumption due to the absence of reconfigurable optics make the developed platform highly promising for end-user and satellite-based quantum photonic systems.

\subsection* {Disclosures} The authors declare no conflicts of interest.

\subsection* {Supplemental document}
See Supplement~1 for supporting content. 

\subsection* {Code, Data, and Materials Availability} 
The code and data that support the findings of this study are available from the authors upon reasonable request.

\subsection* {Acknowledgments}
This work was supported by the Australian Research Council (DP190101559, CE200100010). This work was performed in part at the Melbourne Centre for Nanofabrication (MCN) in the Victorian Node of the Australian National Fabrication Facility (ANFF).


\bibliography{Bibliography}   
\bibliographystyle{spiejour}   





\end{spacing}
\end{document}